\documentclass[superscriptaddress,twocolumn,preprintnumbers,amsmath,amssymb,floatfix,nofootinbib]{revtex4}

\usepackage{graphicx}
\usepackage{dcolumn}
\usepackage{bm}
\usepackage{xcolor}
\usepackage{subfigure}
\usepackage{enumitem}
\usepackage{ulem}
\usepackage[title]{appendix}
\usepackage{array}
\usepackage{lipsum}

\begin{document}

\preprint{}

\title{Impact of resonance decays on critical point signals in net-proton fluctuations}

\author{Marcus Bluhm}
\email{mbluhm@ncsu.edu}
\affiliation{Department of Physics, North Carolina State University, Raleigh, NC 27695, USA}

\author{Marlene Nahrgang}
\affiliation{SUBATECH, UMR 6457, Universit\'e de Nantes, Ecole des Mines de Nantes,
IN2P3/CNRS. 4 rue Alfred Kastler, 44307 Nantes cedex 3, France}
\affiliation{Department of Physics, Duke University, Durham, NC 27708, USA}

\author{Steffen A.~Bass}
\affiliation{Department of Physics, Duke University, Durham, NC 27708, USA}
    
\author{Thomas Sch\"afer}
\affiliation{Department of Physics, North Carolina State University, Raleigh, NC 27695, USA}

%%%%%%%%%%%%%%
%\date{\today}
%%%%%%%%%%%%%%

%%%%%%%%%%%%%%%%%%%%%%%%%%%%%%%%%%%%%%%%%%%%%%%%%%%%%%%%%%%%%%%%%%%%%%%%%%%%%%%%%%%%%%%%%%%%%%%%%%%%%%%%%%%%%%%%%%
\begin{abstract}
The non-monotonic beam energy dependence of the higher cumulants of net-proton fluctuations is a widely studied
signature of the conjectured presence of a critical point in the QCD phase diagram. In this work we study 
the effect of resonance decays on critical fluctuations. We show that resonance effects reduce the signatures
of critical fluctuations, but that for reasonable parameter choices critical effects in the net-proton cumulants 
survive. The relative role of resonance decays has a weak dependence on the order of the cumulants studied with 
a slightly stronger suppression of critical effects for higher-order cumulants. 
\end{abstract}
%%%%%%%%%%%%%%%%%%%%%%%%%%%%%%%%%%%%%%%%%%%%%%%%%%%%%%%%%%%%%%%%%%%%%%%%%%%%%%%%%%%%%%%%%%%%%%%%%%%%%%%%%%%%%%%%%%

%%%%%%%%
%\pacs{}
%%%%%%%%

\maketitle

%%%%%%%%%%%%%%%%%%%%%%%%%%%%%%%%%%%%%%%
\section{Introduction\label{sec:Intro}}
%%%%%%%%%%%%%%%%%%%%%%%%%%%%%%%%%%%%%%%

Hot and dense strongly interacting matter is studied experimentally in heavy-ion collisions at different beam 
energies $\sqrt{s}$. One of the major scientific goals of these experiments is to explore the phases of QCD matter 
as a function of temperature $T$ and baryon chemical potential $\mu_B$. At vanishing $\mu_B$ the transition 
between the high-$T$ partonic and low-$T$ hadronic phases is an analytic crossover~\cite{Aoki:2006we}, but many 
models predict that the transition is first order at large 
$\mu_B$~\cite{Asakawa:1989bq,Berges:1998rc,Halasz:1998qr}. Experimental searches for the first-order phase 
transition and the associated critical endpoint are based on the observation that the baryon chemical potential at 
chemical freeze-out increases with decreasing $\sqrt{s}$ in the regime probed at the Relativistic Heavy Ion 
Collider (RHIC) and the SPS fixed target program at CERN. Promising observable signatures for the presence of a 
critical point include event-by-event fluctuations~\cite{Stephanov:1998dy,Stephanov:1999zu,Hatta:2003wn}, in 
particular fluctuations of conserved quantum numbers such as baryon number or electric 
charge~\cite{Koch:2008ia,Asakawa:2015ybt}. The influence of a phase transition, or a change of its order, should 
be reflected in a non-monotonic $\sqrt{s}$-dependence of these 
observables~\cite{Stephanov:2008qz,Asakawa:2009aj,Athanasiou:2010kw,Friman:2011pf,Stephanov:2011pb,Chen:2016sxn}. 
Moreover, non-uniform structures in e.g.~the net-baryon number distributions due to domain formation in the 
spinodal region of the phase diagram may be expected~\cite{Steinheimer:2012gc,Herold:2013qda}. 

Motivated by these ideas a substantial experimental effort has been directed at studying event-by-event 
multiplicity fluctuations of identified particle species~\cite{Adamczyk:2013dal,Adamczyk:2014fia,Adare:2015aqk}. 
Considered as a proxy for net-baryon number fluctuations, net-proton number fluctuations are particularly 
interesting. For these, intriguing non-monotonic patterns were observed within phase~I of the RHIC beam energy 
scan (BES) program as $\sqrt{s}$ was decreased~\cite{Luo:2015ewa,Luo:2015doi,Thader:2016gpa}. To deduce from the 
observed behavior unambiguously the presence of a first-order phase transition or a critical point is, 
nonetheless, a non-trivial task which requires additional work. On the experimental side the main issue is to 
reduce systematic and statistical uncertainties, which is one of the central goals of the upcoming phase~II of the 
BES. Moreover, the strong sensitivity of the current data on reconstruction efficiency, kinematic acceptance or 
collision-centrality needs to be better understood. On the theoretical side a list of challenges has to be 
resolved before the data can be interpreted meaningfully. For example, net-proton number fluctuations serve, at 
best, as a proxy for net-baryon number fluctuations. In particular at lower $\sqrt{s}$, aspects of exact charge 
conservation~\cite{Schuster:2009jv,Bzdak:2012an}, repulsive interactions~\cite{Fu:2013gga} or late stage 
hadronic phase processes such as resonance regeneration~\cite{Kitazawa:2011wh,Kitazawa:2012at} or decay have 
to be considered carefully. Further details and more fluctuation sources were discussed in the recent 
review~\cite{Nahrgang:2016ayr}. 

The present work addresses one of the theoretical challenges to data interpretation, namely the role of resonance 
decays. Resonance decays contribute significantly to the final state particle multiplicities and, consequently, to 
event-by-event multiplicity distributions. Their impact on net-proton number fluctuations, ignoring the possible 
presence of a critical point, was previously studied in~\cite{Garg:2013ata,Nahrgang:2014fza}. It was found that 
the decay processes do not significantly modify the primordial (without resonance decays) ratios of net-proton 
number cumulants if the probabilistic decay contributions are properly taken into 
account~\cite{Nahrgang:2014fza}. This is because a thermal distribution of particles is being folded with 
a binomial decay distribution yielding again a nearly thermal distribution. However, critical fluctuations 
lead to sizeable deviations from the thermal baseline. In~\cite{Kitazawa:2012at} it was argued that the 
isospin randomization through resonance regeneration processes suppresses strongly the critical fluctuation 
signals in the net-proton number distribution, in particular, in the higher-order cumulants. It is natural to ask 
if resonance decays similarly obscure critical point signals in fluctuation observables. 

Critical fluctuations can be described in terms of an order parameter field $\sigma$. In the thermodynamic 
limit the equilibrium correlation length $\xi$ of the order parameter diverges at the critical point. Under 
the assumption that QCD belongs to the same universality class as the three-dimensional Ising 
model~\cite{Berges:1998rc,Halasz:1998qr} one finds that the cumulants of the order parameter fluctuations 
scale as $\langle (\delta\sigma)^2 \rangle \propto \xi^2$, $\langle (\delta\sigma)^3 \rangle \propto \xi^{9/2}$ 
and $\langle (\delta\sigma)^4 \rangle_c \propto \xi^7$, where small anomalous dimension corrections are neglected 
in these scaling relations~\cite{Stephanov:2008qz}. Thus, higher-order cumulants depend more sensitively on the 
value of the equilibrium correlation length. 

This is important for heavy-ion collisions, because the actual value of $\xi$ is limited due to finite size and 
time effects. Most significantly, the temporal growth of the correlation length is controlled by the relaxation 
time $\tau_\xi$, which increases as a power of the equilibrium correlation length. Near the critical point 
$\tau_\xi\sim\xi^z$ with $z\approx 3$~\cite{Son:2004iv}, and critical slowing down places significant limits on 
the actual growth of the correlation length. In~\cite{Berdnikov:1999ph}, the non-equilibrium evolution of the 
correlation length was studied and a maximal value of about $2-3$~fm was found. Thus, critical fluctuation signals 
in observables will be weaker than expected based on equilibrium considerations. Within the model of 
non-equilibrium chiral fluid dynamics~\cite{Nahrgang:2011mg,Herold:2013bi} fluctuations of the order parameter are 
explicitly propagated in a fluid dynamical evolution of heavy-ion collisions. The effect of such a dynamics on the 
time evolution of net-proton fluctuations has been investigated recently~\cite{Herold:2016uvv}. It was found that 
critical fluctuations are affected by the dynamics, and that the location on the hydrodynamic trajectory where 
fluctuations freeze out determines the magnitude and shape of the fluctuation signals. However, it was also found 
that a critical point signal remains visible on top of thermal and initial state fluctuations. An analytic model 
for the time evolution of order parameter fluctuations was recently studied in~\cite{Mukherjee:2015swa}. The 
authors found that relaxation time effects introduce a significant lag in the cumulants relative to equilibrium 
expectations, and that non-equilibrium effects modify the simple scaling relations discussed 
in~\cite{Stephanov:2008qz}. 

In this work we assume that there is a critical point in the QCD phase diagram, that this point causes critical 
fluctuations in the primordial net-proton number at chemical freeze-out, and that these fluctuations are 
determined by the growth of the correlation length. We study quantitatively the impact of resonance decays on the 
higher-order net-proton number cumulant ratios. We do not take into account dynamical effects such as the role of 
the relaxation time and critical slowing down. We assume that a) the fluctuations reflect thermal equilibrium at 
chemical freeze-out, and that b) freeze-out takes place in proximity to the QCD phase transition. These 
assumptions are motivated by the success of fluid dynamics, based on the assumption of rapid local thermalization, 
in describing bulk observables~\cite{Gale:2013da,Bernhard:2016tnd}, and by the successful extraction of freeze-out 
parameters from fluctuation observables~\cite{Alba:2014eba}. Certainly, results obtained in the limit of rapid 
local thermalization provide an important benchmark for more sophisticated studies based on non-equilibrium 
thermodynamics. 

%%%%%%%%%%%%%%%%%%%%%%%%%%%%%%%%%%%%%%%%%%%%%%%%%%%%%%%%%%%%%%%%%%%%%%%%%%%%%%%%%%%%%
\section{Net-proton fluctuations and the impact of a critical point\label{sec:Model}}
%%%%%%%%%%%%%%%%%%%%%%%%%%%%%%%%%%%%%%%%%%%%%%%%%%%%%%%%%%%%%%%%%%%%%%%%%%%%%%%%%%%%%

In this section, we discuss in detail the framework used in our work. We start by describing the thermal, 
i.e.~non-critical, baseline which we assume to be given by a grandcanonical hadron resonance gas model. 
It was shown that such a model is quite successful in quantitatively describing both lattice QCD results of the 
thermodynamics in the hadronic phase~\cite{Karsch:2003vd,Bellwied:2015lba} and measured particle 
yields~\cite{Andronic:2011yq}. Critical fluctuations are then included by making use of universality class 
arguments and coupling (anti-)particles with the order parameter field, the critical mode, via a phenomenological 
approach. We discuss how the critical fluctuations can be quantified at the chemical freeze-out and, furthermore, 
propose different phenomenological ans\"atze to study, in addition, the impact resonance decays have on the 
net-proton number fluctuations in the presence of a critical point. 

%%%%%%%%%%%%%%%%%%%%%%%%%%%%%%%%%%%%%%%%%%%%%%%%%%%%%%
\subsection{Non-critical baseline model\label{sec:2A}}
%%%%%%%%%%%%%%%%%%%%%%%%%%%%%%%%%%%%%%%%%%%%%%%%%%%%%%

As a baseline theory that exhibits no critical behavior we consider a hadron resonance gas model containing 
baryons and mesons up to masses of approximately $2$~GeV in line with~\cite{Beringer:1900zz}. Starting point for 
our study is the particle density of species $i$ given by the momentum-integral 
\begin{equation}
 \label{equ:particledensity}
 n_i(T,\mu_i)=d_i \int \frac{{\rm d}^3k}{(2\pi)^3}\, f_i^0(T,\mu_i)
\end{equation}
with degeneracy factor $d_i$ and distribution function 
\begin{equation}
 \label{equ:distfunction}
 f_i^0(T,\mu_i)=\frac{1}{(-1)^{B_i+1}+e^{(E_i-\mu_i)/T}} \,.
\end{equation}
In Eq.~(\ref{equ:distfunction}), $E_i=\sqrt{k^2+m_i^2}$ is the energy of particles of species $i$ with mass 
$m_i$, and $\mu_i=B_i\mu_B+S_i\mu_S+Q_i\mu_Q$ is their chemical potential, where $\mu_X$ are conjugate to the 
conserved charges $N_X$ and $B_i,\, S_i,\, Q_i$ represent the corresponding quantum numbers of baryon number, 
strangeness and electric charge, respectively. 

For the grandcanonical ensemble of a non-interacting hadron resonance gas, one defines the mean particle 
number $\langle N_i\rangle$ of species $i$ in a fixed volume $V$ as $\langle N_i\rangle=Vn_i$. The cumulants 
with $n>1$, measuring thermal ensemble fluctuations in the particle number, can be related to derivatives of 
$n_i/T^3$ with respect to $\mu_i/T$ for fixed $T$, 
\begin{equation}
 \label{equ:CnDefinition}
 C_n^i = VT^3 \left.\frac{\partial^{n-1}(n_i/T^3)}{\partial(\mu_i/T)^{n-1}}\right|_T\,.
\end{equation}
In this work, we are interested in the first four cumulants $C_n$ of the net-proton number 
$N_{p-\bar{p}}=N_p-N_{\bar{p}}$, which read (note that cumulant and central moment are the same for order 
$n\leq 3$) 
\begin{align}
 \label{equ:C1Def}
 C_1 & = \langle N_{p-\bar{p}}\rangle =  \langle N_p\rangle - \langle N_{\bar{p}}\rangle \,,\\
 \label{equ:C2Def}
 C_2 & = \langle (\Delta N_{p-\bar{p}})^2\rangle = C_2^p+C_2^{\bar{p}}\,,\\
 \label{equ:C3Def}
 C_3 & = \langle (\Delta N_{p-\bar{p}})^3\rangle = C_3^p-C_3^{\bar{p}}\,,\\
 \label{equ:C4Def}
 C_4 & = \langle (\Delta N_{p-\bar{p}})^4\rangle_c = C_4^p+C_4^{\bar{p}}
\end{align}
with $\langle (\Delta N_{p-\bar{p}})^4\rangle_c = \langle (\Delta N_{p-\bar{p}})^4\rangle 
- 3\!\left.\langle (\Delta N_{p-\bar{p}})^2\rangle\right.^{\!\!2}$ and 
$\Delta N_{p-\bar{p}}=N_{p-\bar{p}}-\langle N_{p-\bar{p}}\rangle$. The second equalities in 
Eqs.~(\ref{equ:C2Def})~-~(\ref{equ:C4Def}) are specific for our baseline model, for which correlations among 
different particle species vanish. 

It is useful to consider particular ratios of these cumulants. Assuming that the fluctuations originate from a 
source with fixed thermal conditions, we may define 
\begin{equation}
 \label{equ:Cnratios}
 \frac{C_2}{C_1} = \frac{\sigma^2}{M}\,, \quad \frac{C_3}{C_2} = S\sigma\,, \quad 
 \frac{C_4}{C_2} = \kappa\sigma^2 
\end{equation}
and compare model calculations for the cumulant ratios with combinations of mean $M=C_1$, variance $\sigma^2=C_2$, 
skewness $S=C_3/C_2^{3/2}$ and kurtosis $\kappa=C_4/C_2^2$ of measured event-by-event multiplicity distributions. 
An advantage of considering ratios is that to first approximation volume fluctuations, which are inherent in the 
individual cumulants $C_n$, cancel out. 

%%%%%%%%%%%%%%%%%%%%%%%%%%%%%%%%%%%%%%%%%%%%%%%%%%%%%%%%%%%%%%%%%%%%%%%%%%%%%%%%%%%%%%%%%%
\subsection{Critical mode fluctuations and coupling to physical observables\label{sec:2B}}
%%%%%%%%%%%%%%%%%%%%%%%%%%%%%%%%%%%%%%%%%%%%%%%%%%%%%%%%%%%%%%%%%%%%%%%%%%%%%%%%%%%%%%%%%%

In the scaling region near a critical point different physical systems, which belong to the same universality 
class, exhibit a universal critical behavior~\cite{Hohenberg:1977ym}. Assuming that QCD belongs to the same 
universality class as the three-dimensional Ising model allows us to relate the order parameter $\sigma$ of the 
chiral phase transition in QCD with the order parameter of the spin model, the magnetization $M$. In the Ising 
model, the magnetization is a function of reduced temperature $r=(T-T_c)/T_c$ and reduced external magnetic field 
$h=H/H_0$, where $T_c$ is the critical temperature in the spin model and $H_0$ is a normalization constant. In 
these variables, the critical point is located at $r=h=0$. For $r<0$ and $h=0$ the transition is of first order, 
while $r>0$ marks the crossover regime. 

The equilibrium cumulants of the critical mode are then obtained as 
\begin{equation}
\label{equ:sigmacumulants}
 \langle(\delta\sigma)^n\rangle_c = 
 \left(\frac{T}{VH_0}\right)^{n-1}\left.\frac{\partial^{n-1}M}{\partial h^{n-1}}\right|_r 
\end{equation}
from appropriate derivatives of $M$ with respect to the classical field keeping $r$ 
fixed~\cite{Stephanov:2011pb,Mukherjee:2015swa,Chen:2016sxn}. A parametric representation of $M$ based 
on~\cite{Guida:1996ep} can be found in Appendix~\ref{sec:AppA}. This gives rise to parametric expressions for 
the cumulants of the critical mode, which are also given in Appendix~\ref{sec:AppA}. These expressions are, 
strictly speaking, only valid for the scaling region close to the critical point. Nevertheless, we employ them 
for all considered $\sqrt{s}$ because a) the size of the scaling region is not known, and b) the influence of 
critical fluctuations on observables is suppressed further away from the critical point as we discuss in 
section~\ref{sec:Results}. 

Fluctuations of the critical mode are not directly measurable. Instead, we focus on observables that couple to the 
critical mode, such as the number of pions or protons~\cite{Stephanov:1999zu,Hatta:2003wn}. In general, the 
coupling of these observables to the critical mode has to be modeled. In the following we will take the order 
parameter to be the chiral field $\sigma$, and model the coupling of (anti-)particles to the critical mode in a 
way that is suggested by models in which the chiral field has a linear relation to the dynamically generated mass, 
$\delta m_i=g_i\delta\sigma$~\cite{Stephanov:2011pb,Athanasiou:2010kw,Stephanov:1999zu}. As a consequence, 
fluctuations 
\begin{equation}
 \label{equ:DeltaDist}
 \delta f_i = - \delta\sigma \frac{g_i}{T} \frac{v_i^2}{\gamma_i} 
\end{equation}
associated with the variations in the particle masses arise in addition to the thermal ensemble fluctuations 
discussed in section~\ref{sec:2A}. In Eq.~(\ref{equ:DeltaDist}), $\gamma_i=E_i/m_i$, 
$v_i^2=f_i^0\left((-1)^{B_i}f_i^0+1\right)$ and $\langle\delta\sigma\rangle=0$ such that 
$\langle\delta m_i\rangle=0$ on average. 

The coupling of protons and anti-protons to the critical mode with coupling strength $g_p$ influences the 
net-proton number fluctuations near the critical point. According to Eq.~(\ref{equ:DeltaDist}), critical 
fluctuations do not affect the mean value $C_1$ but modify the variance 
$C_2=\langle (\Delta N_p)^2 \rangle + \langle (\Delta N_{\bar{p}})^2 \rangle - 
2 \langle\Delta N_p\Delta N_{\bar{p}} \rangle$ via 
\begin{align}
 \label{equ:C2crit}
 \langle(\Delta N_i)^2 \rangle & = C_2^i + \langle(V\delta\sigma)^2\rangle I_i^2 \,, \\
 \label{equ:C11crit}
 \langle\Delta N_i\Delta N_j \rangle & = \langle(V\delta\sigma)^2\rangle I_i I_j 
\end{align}
with 
\begin{equation}
 \label{equ:IFunction}
 I_i = \frac{g_id_i}{T} \int \frac{d^3k}{(2\pi)^3} \frac{v_i^2}{\gamma_i} \,.
\end{equation}
In these expressions we assumed that non-critical and critical fluctuations are uncorrelated. The coupling to 
the critical mode induces correlations between protons and anti-protons in $C_2$, which reduce the variance 
compared to that in the case of an independent production. Simultaneously, critical fluctuations enhance the 
individual proton and anti-proton variances according to Eq.~(\ref{equ:C2crit}). With the help of 
Eq.~(\ref{equ:IFunction}) we can write $C_2$ in compact form as 
\begin{equation}
 \label{equ:C2compact}
 C_2 = C_2^p + C_2^{\bar{p}} + \langle (V\delta\sigma)^2 \rangle \left(I_p-I_{\bar{p}}\right)^2 
\end{equation}
and find as net-effect an enhancement of the net-proton variance due to critical fluctuations. 

Similarly, $C_3$ and $C_4$ are modified. For $C_3 = \langle(\Delta N_p)^3 \rangle - 
\langle(\Delta N_{\bar{p}})^3 \rangle - 3 \langle(\Delta N_p)^2\Delta N_{\bar{p}} \rangle + 
3 \langle\Delta N_p(\Delta N_{\bar{p}})^2 \rangle$ we find 
\begin{align}
 \label{equ:C3crit}
 \langle(\Delta N_i)^3 \rangle & = C_3^i - \langle(V\delta\sigma)^3\rangle I_i^3 \,, \\
 \label{equ:C21crit}
 \langle(\Delta N_i)^2\Delta N_j \rangle & = - \langle(V\delta\sigma)^3\rangle I_i^2 I_j \,,
\end{align}
where we neglected subdominant critical fluctuation contributions of order lower than ${\cal O}(\xi^{9/2})$. 
Accordingly, we can write 
\begin{equation}
 \label{equ:C3compact}
 C_3 = C_3^p - C_3^{\bar{p}} - \langle (V\delta\sigma)^3 \rangle \left(I_p-I_{\bar{p}}\right)^3 \,.
\end{equation}
For $C_4 = \langle(\Delta N_p)^4 \rangle_c + \langle(\Delta N_{\bar{p}})^4 \rangle_c 
- 4 \langle(\Delta N_p)^3\Delta N_{\bar{p}} \rangle_c + 6 \langle(\Delta N_p)^2(\Delta N_{\bar{p}})^2 
\rangle_c - 4 \langle\Delta N_p(\Delta N_{\bar{p}})^3 \rangle_c$ one has 
\begin{align}
 \label{equ:C4crit}
 \langle(\Delta N_i)^4 \rangle_c & = C_4^i + \langle(V\delta\sigma)^4\rangle_c \,I_i^4 \,, \\
 \label{equ:Cmncrit}
 \langle(\Delta N_i)^m(\Delta N_j)^n \rangle_c & = \langle(V\delta\sigma)^4\rangle_c \,I_i^m I_j^n 
\end{align}
for $m+n=4$, where we also neglected subdominant critical fluctuation contributions, here of order lower than 
${\cal O}(\xi^7)$. In compact form one may write 
\begin{equation}
 \label{equ:C4compact}
 C_4 = C_4^p + C_4^{\bar{p}} + \langle (V\delta\sigma)^4 \rangle_c \left(I_p-I_{\bar{p}}\right)^4 \,.
\end{equation}
We note that if non-critical and critical fluctuations are independent, then Eqs.~(\ref{equ:C3compact}) 
and~(\ref{equ:C4compact}) will represent the full results for $C_3$ and $C_4$, respectively, in line with the 
model assumption Eq.~(\ref{equ:DeltaDist}). 

%%%%%%%%%%%%%%%%%%%%%%%%%%%%%%%%%%%%%%%%%%%%%%%%%%%%%%%%
\subsection{Mapping to QCD thermodynamics\label{sec:2C}}
%%%%%%%%%%%%%%%%%%%%%%%%%%%%%%%%%%%%%%%%%%%%%%%%%%%%%%%%

\begin{figure}[tb]
 \vspace{4mm}
 \includegraphics[width=0.43\textwidth]{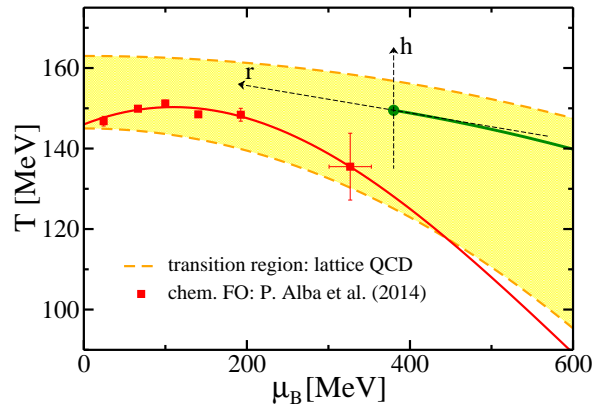}
 \caption{(Color online) Sketch of the set-up in the QCD phase diagram considered in this work. The filled band 
 between the two dashed curves comprises available information from lattice QCD on the crossover region as known 
 for small $\mu_B$, which we extrapolate to larger $\mu_B$. The dashed curves are obtained by solving 
 Eq.~(\ref{equ:crossoverline}) to leading order with $\kappa_c=0.007$ (upper curve, $T_{c,0}=0.163$~GeV) and 
 $\kappa_c=0.02$ (lower curve, $T_{c,0}=0.145$~GeV). The location of the critical point and of the adjacent 
 first-order phase transition is not known. The placement in this figure corresponds to the 
 choice of parameters specified in section~\ref{sec:3B}. The spin model coordinate system $(r,h)$ has its 
 origin in the critical point. Details of the mapping are described in 
 the text. We also show the chemical freeze-out conditions determined in~\cite{Alba:2014eba} (squares) and a 
 practical parametrization of these results (solid curve), which is discussed in Appendix~\ref{sec:AppB}.} 
 \label{fig:fig1}
\end{figure}

In order to quantify the effect of the critical point on net-proton number fluctuations, we have to specify the 
cumulants of the critical mode $\langle(\delta\sigma)^n\rangle_c$ at chemical freeze-out. In this work we use
the chemical freeze-out conditions reported in~\cite{Alba:2014eba}, which are shown in Fig.~\ref{fig:fig1} 
(squares). These were determined by analyzing the STAR data~\cite{Adamczyk:2013dal,Adamczyk:2014fia} on net-proton 
number and net-electric charge fluctuations. A discussion and practical parametrization of the results (exhibited 
by the solid curve in Fig.~\ref{fig:fig1}) is relegated to Appendix~\ref{sec:AppB}. 
%%%%%%%%%%%%%%%%%%%%%%%%%%%%%%%%%%%%%%%%%%%%%%%%%%%%%%%%%%%%%%%%%%%%%%%%%%%%%%%%%%%%%%%%%%%%%%%%%%%%%%%%%%%%%%%%%
%%%%%%%%%%%%%%%%%%%%%%%%%%%%%%%%%%%%%%%%%%%%%%%%%%%%%%%%%%%%%%%%%%%%%%%%%%%%%%%%%%%%%%%%%%%%%%%%%%%%%%%%%%%%%%%%%

For any set of values of Ising model variables $r$ and $h$, we can determine the cumulants 
$\langle(\delta\sigma)^n\rangle_c$ from the parametric representation discussed in detail in 
Appendix~\ref{sec:AppA} by using the transformations in Eqs.~(\ref{equ:A2}) and~(\ref{equ:A3}). Therefore, we 
only have to map the thermal variables in QCD, $T$ and $\mu_B$, to the variables of the spin model. This mapping 
is non-universal: The relation between $r$ and $h$ on the one hand, and $T$ and $\mu_B$ on the 
other, depends sensitively on model assumptions. The easiest approach, used frequently in the 
literature~\cite{Mukherjee:2015swa,Nonaka:2004pg}, is a linear map $r(T,\mu_B)$ and $h(T,\mu_B)$ subject 
to two conditions: a) the conjectured QCD critical point $(\mu_B^{\textrm{cp}},T^{\textrm{cp}})$ is located 
in Ising coordinates at $r=h=0$, and b) the $r$-axis must lie tangentially to the first-order phase 
transition line in the QCD critical point and be oriented such that positive $r$-values correspond 
to the crossover regime in QCD (see sketch in Fig.~\ref{fig:fig1}). 

The exact location of the critical point and the slope of the adjacent first-order phase transition line are not 
known. In this work we use lattice information on the location of the pseudo-critical line in the 
$(\mu_B,T)$-plane, indicated by the filled band between the two dashed curves in Fig.~\ref{fig:fig1}. The lattice 
data can be parameterized as 
\begin{equation}
 \label{equ:crossoverline}
 T_c(\mu_B)=T_{c,0}\left[1-\kappa_c\left(\frac{\mu_B}{T_c(\mu_B)}\right)^2+\dots\right] \,,
\end{equation}
where $T_{c,0}=(0.145\dots 0.163)$~GeV is the pseudo-critical temperature at vanishing 
$\mu_B$~\cite{Borsanyi:2010bp,Bazavov:2011nk} and $\kappa_c\simeq 0.007\dots 0.059$ is the chiral crossover 
curvature~\cite{Kaczmarek:2011zz,Endrodi:2011gv,Bonati:2015bha,Bellwied:2015rza,Cea:2015cya}. 

The mapping to $h$ is not well constrained. A typical choice made in the literature is to define the $h$-axis 
perpendicular to the $r$-axis~\cite{Mukherjee:2015swa,Nonaka:2004pg}. However, since the first-order phase 
transition line is expected to have some curvature in the $(\mu_B,T)$-plane it is clear that, in 
general, the map must be more complicated. In this work, we follow~\cite{Mukherjee:2015swa} and define 
\begin{equation}
 \label{equ:hdefinition}
 h = \frac{(T-T^{\textrm{cp}})}{\Delta T^{\textrm{cp}}}
\end{equation}
such that the $h$-axis is simply parallel to the $T$-axis (see Fig.~\ref{fig:fig1}). We also define
the auxiliary variable 
\begin{equation}
 \label{equ:rdefinition}
 \tilde{r} = \frac{(\mu_B-\mu_B^{\textrm{cp}})}{\Delta \mu_B^{\textrm{cp}}} \,.
\end{equation}
The parameters $\Delta T^{\textrm{cp}}$ and $\Delta \mu_B^{\textrm{cp}}$ determine the size of the 
critical region. The Ising variable $r$ is then obtained by rotating $\tilde{r}$, where the slope of the $r$-axis 
is determined by using the slope of $T_c(\mu_B)$ at $\mu_B^{\textrm{cp}}$. 

%%%%%%%%%%%%%%%%%%%%%%%%%%%%%%%%%%%%%%%%%%%
\subsection{Resonance decays\label{sec:2D}}
%%%%%%%%%%%%%%%%%%%%%%%%%%%%%%%%%%%%%%%%%%%

The cumulants of a multiplicity distribution influenced by resonance decays were derived 
in~\cite{Fu:2013gga,Begun:2006jf}. These results take the probabilistic nature of the decay process into account. 
As critical fluctuations do not alter mean values, cf.~Eq.~(\ref{equ:DeltaDist}), the mean of the final net-proton 
number reads 
\begin{equation}
 \label{equ:C1critReso}
 \hat{C}_1 = \langle N_p\rangle - \langle N_{\bar{p}}\rangle + \sum_R \langle N_R\rangle 
 \left(\langle n_p\rangle_R - \langle n_{\bar{p}}\rangle_R\right) \,,
\end{equation}
where the notation $\hat{C}_n$ refers to a cumulant that includes resonance decays. We also define 
$\langle n_i\rangle_R=\sum_r b_r^R n_{i,r}^R$ as the decay-average number of particles of species $i$ created from 
resonance $R$, where $n_{i,r}^R$ are created in decay branch $r$ with branching ratio $b_r^R$. Note that the sum 
in Eq.~(\ref{equ:C1critReso}) includes all resonances and anti-resonances of the baseline hadron resonance gas 
model. Analogous formulas for the higher cumulants, which contain critical fluctuation contributions, are given in 
Appendix~\ref{sec:AppC}. 

To study the impact of resonance decays on the higher-order cumulants of the net-proton number in the presence of 
a critical point we must quantify the coupling strengths $g_R$ between the critical mode and the 
(anti-)resonances. In line with the chiral model study~\cite{Zschiesche:2000ew} we may assume that $g_R$ is 
proportional to $g_p$ for non-strange baryonic resonances like $\Delta^{++}$ with a proportionality factor that 
reflects the mass difference between the resonance and the proton. For strangeness carrying resonances like 
$\Lambda(1520)$ we must, in addition, account for the fact that their mass generation is less dominated by the 
coupling to the chiral field~\cite{Zschiesche:2000ew}. To accommodate these features we advocate as a 
phenomenological ansatz 
\begin{equation}
 \label{equ:sigmaresocoupling}
 g_R=\frac{g_p}{3}\frac{m_R}{m_p}(3-|S_R|) 
\end{equation}
for the $\sigma$-(anti-)resonance couplings. To analyze the impact of other possible scenarios, we also consider 
the two cases that either all resonances couple like protons, $g_R=g_p$, or that resonances do not couple to the 
critical mode at all, $g_R=0$. This allows us to study how sensitively the final results depend on the choice for 
the couplings $g_R$. 

%%%%%%%%%%%%%%%%%%%%%%%%%%%%%%%%%%%%%%%%%%%%%%
\section{Numerical results\label{sec:Results}}
%%%%%%%%%%%%%%%%%%%%%%%%%%%%%%%%%%%%%%%%%%%%%%

In this section we present numerical results of our study. We start by discussing how to implement kinematic cuts 
in order to take into account the experimental acceptance. Then, we quantify the effect of resonance decays on 
critical point signatures in net-proton number fluctuations. We compare our results to preliminary STAR 
data~\cite{Thader:2016gpa}. We should note that there are sizeable uncertainties in both the data and the 
theoretical analysis. Our goal in this work is not to establish or constrain the existence of a critical point, 
but to quantify the role of resonance decays. 

%%%%%%%%%%%%%%%%%%%%%%%%%%%%%%%%%%%%%%%%% 
\subsection{Kinematic cuts\label{sec:3A}}
%%%%%%%%%%%%%%%%%%%%%%%%%%%%%%%%%%%%%%%%%

In the analyses~\cite{Adamczyk:2013dal,Luo:2015ewa,Luo:2015doi,Thader:2016gpa} of net-proton number fluctuations 
the detector acceptance was limited. Coverage in kinematic rapidity was restricted to the mid-rapidity region 
$|y|\leq 0.5$, with a full coverage in azimuthal angle $\phi$, and in transverse momentum cuts of 
$0.4~{\rm GeV}\leq k_T\leq k_T^{\rm max}$ with $k_T^{\rm max}=0.8$~GeV~\cite{Adamczyk:2013dal} or 
$k_T^{\rm max}=2$~GeV~\cite{Luo:2015ewa,Luo:2015doi,Thader:2016gpa} were applied. For a smaller experimentally 
analyzed phase space, one expects critical fluctuation signals in observables to be diminished because the 
momentum-space correlations, which are caused by the potentially large spatial correlations through thermal 
smearing, may exceed the considered kinematic acceptance window~\cite{Ling:2015yau,Ohnishi:2016bdf}. Indeed, the 
pronounced structures seen in the preliminary data of the higher-order net-proton number cumulant 
ratios~\cite{Luo:2015ewa,Luo:2015doi,Thader:2016gpa} disappear to a large extent in the published 
data~\cite{Adamczyk:2013dal} for smaller $k_T^{\rm max}$. 

In this work, we include the kinematic cuts on the level of the momentum-integrals, 
cf.~Eqs.~(\ref{equ:particledensity}) and~(\ref{equ:IFunction}). For this purpose, we rewrite 
following~\cite{Garg:2013ata} 
\begin{equation}
 \int {\rm d}^3k \,\to\, \int k_T\sqrt{k_T^2+m_i^2}\cosh(y) \,{\rm d}k_T \,{\rm d}y \,{\rm d}\phi \,,
\end{equation}
limit the integrations to $-0.5\leq y\leq 0.5$, $0\leq\phi\leq 2\pi$ and $0.4~{\rm GeV}\leq k_T\leq k_T^{\rm max}$ 
and replace $E_i$ by $\cosh(y)\sqrt{k_T^2+m_i^2}$. This implies that we apply the kinematic cuts to evaluate 
quantities at the chemical freeze-out and neglect the subsequent modifications of particle multiplicity 
distributions in a given acceptance window due to elastic scatterings until kinetic freeze-out. 

%%%%%%%%%%%%%%%%%%%%%%%%%%%%%%%%%%%%%%%%%%
\subsection{Cumulant ratios\label{sec:3B}}
%%%%%%%%%%%%%%%%%%%%%%%%%%%%%%%%%%%%%%%%%%

Before we can show our results we have to fix the parameters that enter into our calculation. These include the 
pseudo-critical line parameters $T_{c,0}$ and $\kappa_c$, the coordinates of the critical point 
$(\mu_B^{\textrm{cp}},T^{\textrm{cp}})$, the width of the critical regime $(\Delta\mu_B^{\textrm{cp}}, 
\Delta T^{\textrm{cp}})$, the normalizations of the critical equation of state $M_0$ and $H_0$, and the coupling 
strength $g_p$. 

We assume $T_{c,0}=0.156$~GeV and $\kappa_c=0.007$ based on the lattice data cited in section~\ref{sec:2C}. The 
STAR data show tentative hints of critical behavior in the kurtosis in the regime $\sqrt{s}=(14.5-19.6)$~GeV. We 
choose $\mu_B^{\textrm{cp}}=0.39$~GeV to illustrate the effects of possible critical behavior in this regime. The 
pseudo-critical line then predicts $T^{\textrm{cp}}\simeq 0.149$~GeV. We use $\Delta T^{\textrm{cp}}=0.02$~GeV and 
$\Delta\mu_B^{\textrm{cp}}=0.42$~GeV, and set the spin model normalization constants to 
$M_0/$GeV$=5.52\cdot 10^{-2}$ and $H_0/$GeV$^3=3.44\cdot 10^{-4}$. The resulting behavior of the correlation 
length along the chemical freeze-out curve is shown in Fig.~\ref{fig:fig5} (see Appendix~\ref{sec:AppA}). We 
observe that the maximum of the correlation length is about 2~fm, comparable to estimates based on dynamical 
models. 

To quantify the influence of critical fluctuations on the net-proton number cumulant ratios, we also need to 
specify the value of the coupling strength $g_p$ between the (anti-)protons and the critical mode, which enters 
the cumulants via the factors $I_p$ and $I_{\bar{p}}$, cf.~Eq.~(\ref{equ:IFunction}). In the context of the 
linear sigma model the coupling constant $g_p$ in the ground state can be related to the pion decay 
constant~\cite{Stephanov:2008qz,Hatta:2003wn}, $g_p\simeq m_p/f_\pi\simeq 10$. In chiral models, parameters are 
typically fixed to reproduce the vacuum masses of baryons and mesons as well as other known properties of nuclear 
matter at saturation density. In the non-linear chiral model considered in~\cite{Dexheimer:2008ax} to describe QCD 
matter in neutron stars, for example, $|g_p|\lesssim 10$ was used. Quark model calculations~\cite{Downum:2006re} 
of nucleon-meson couplings find $g_p$ in the range of $3$~-~$7$, instead. To illustrate the importance of the 
actual value of $g_p$, we use three different values $g_p=3,\,5$ and $7$ in the following. 

\begin{figure}[tb]
 \includegraphics[width=0.47\textwidth]{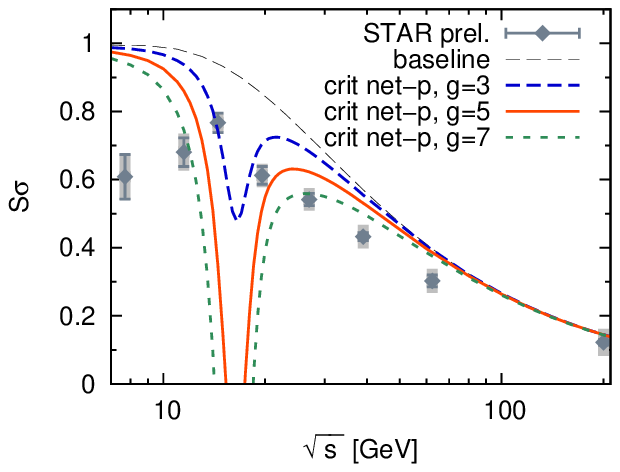}\\ \vspace{2mm}
 \includegraphics[width=0.47\textwidth]{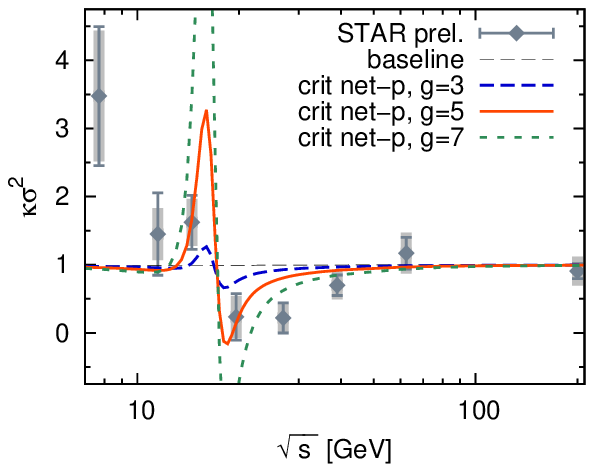}
 \caption{(Color online) Cumulant ratios $S\sigma$ and $\kappa\sigma^2$ for net-protons as functions of $\sqrt{s}$ 
 without including resonance decays. Shown results are based on Eqs.~(\ref{equ:C2Def})~-~(\ref{equ:C4Def}) as 
 baseline (thin dashed curves) without critical fluctuations, and based on Eqs.~(\ref{equ:C2compact}), 
 (\ref{equ:C3compact}) and~(\ref{equ:C4compact}) including critical fluctuations for the exemplary 
 $\sigma$-(anti-)proton couplings $g=g_p=3,\,5,\,7$. These results are obtained along the chemical freeze-out 
 curve shown in Fig.~\ref{fig:fig1} and include the kinematic cuts applied in~\cite{Thader:2016gpa}. For guidance, 
 the preliminary data~\cite{Thader:2016gpa} are also shown. See text for more details.} 
 \label{fig:fig2}
\end{figure}

In Fig.~\ref{fig:fig2}, we show the net-proton number cumulant ratios $S\sigma$ and $\kappa\sigma^2$ defined in 
Eq.~(\ref{equ:Cnratios}) without including the contributions from resonance decays. The non-critical baseline 
(thin dashed curves) is determined using Eqs.~(\ref{equ:C2Def})~-~(\ref{equ:C4Def}). The relevant expressions, 
cf.~Eq.~(\ref{equ:CnDefinition}), are evaluated along a chemical freeze-out curve, where we employ the 
$\sqrt{s}$-parametrizations of the thermal variables $T^{\textrm{fo}}$ and $\mu_X^{\textrm{fo}}$ summarized in 
Appendix~\ref{sec:AppB}. Kinematic cuts are applied as described in section~\ref{sec:3A}, where we use 
$k_T^{\rm max}=2$~GeV in line with~\cite{Thader:2016gpa}. To include critical fluctuation contributions we use 
Eqs.~(\ref{equ:C2compact}), (\ref{equ:C3compact}) and~(\ref{equ:C4compact}). The relevant cumulants of the 
critical mode $\langle(\delta\sigma)^n\rangle_c$ are obtained via Eqs.~(\ref{equ:sigmaC2})~-~(\ref{equ:sigmaC4}) 
by mapping $T$ and $\mu_B$ at chemical freeze-out to the Ising variables $r$ and $h$ as explained in 
section~\ref{sec:2C}. 
%%%%%%%%%%%%%%%%%%%%%%%%%%%%%%%%%%%%%%%%%%%%%%%%%%%%%%%%%%%%%%%%%%%%%%%%%%%%%%%%%%%%%%%%%%%%%%%%%%%%%%%%%%%%%%%%%%
%%%%%%%%%%%%%%%%%%%%%%%%%%%%%%%%%%%%%%%%%%%%%%%%%%%%%%%%%%%%%%%%%%%%%%%%%%%%%%%%%%%%%%%%%%%%%%%%%%%%%%%%%%%%%%%%%%

As expected, critical fluctuations cause a non-monotonic behavior of the net-proton number cumulant ratios. The 
qualitative features are driven by the sign changes in $\langle(\delta\sigma)^3\rangle$ and 
$\langle(\delta\sigma)^4\rangle_c$ as discussed in Appendix~\ref{sec:AppA}. The effects are found to be stronger 
for larger $g_p$ and in the region of beam energies, where the correlation length $\xi$ at chemical freeze-out 
is largest, cf. Fig.~\ref{fig:fig5} in Appendix~\ref{sec:AppA}. Further away from this region, at small and 
large $\sqrt{s}$, the contributions from critical fluctuations are suppressed and the net-proton number cumulant 
ratios fall back to the baseline results. At small $\sqrt{s}$, this is a consequence of the smallness of $\xi$ 
further away from the critical point. But more importantly, in particular for large $\sqrt{s}$, this is caused by 
the thermal factor $(I_p-I_{\bar{p}})/g_p$, which is positive definite and a monotonically decreasing function 
with increasing $\sqrt{s}$ along the chemical freeze-out curve. 

\begin{figure}[tb]
 \includegraphics[width=0.47\textwidth]{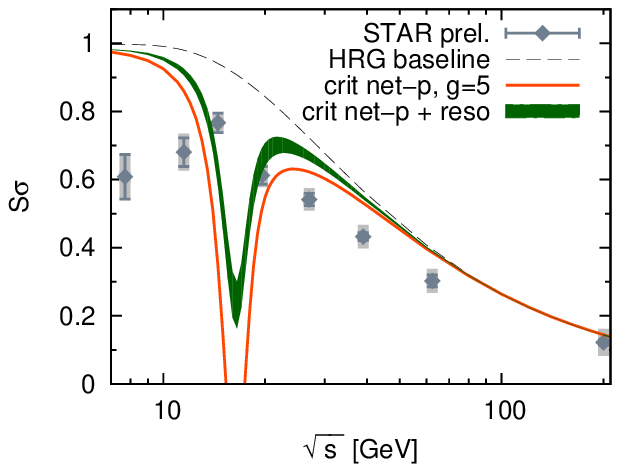}\\ \vspace{2mm}
 \includegraphics[width=0.47\textwidth]{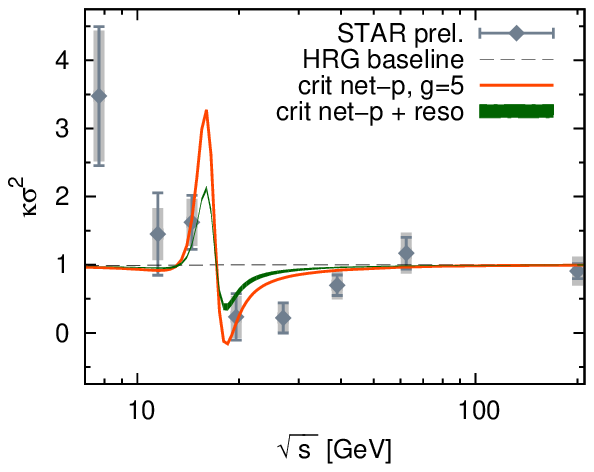}
 \caption{(Color online) Similar to Fig.~\ref{fig:fig2}, but including resonance decays both in the baseline 
 results (thin dashed curves), cf.~\cite{Nahrgang:2014fza} for more details, and for the case $g=g_p=5$. Resonance 
 decays are found to reduce the influence of critical fluctuations on the net-proton number cumulant ratios 
 visibly. The filled bands contain the results of the three different ans\"atze for $g_R$ discussed in this work: 
 the strongest reduction of baseline deviations is seen for $g_R=0$ when resonances do not couple to the critical 
 mode, while critical fluctuations are least suppressed by resonance decays for $g_R$ following the 
 phenomenological ansatz Eq.~(\ref{equ:sigmaresocoupling}). The case $g_R=g_p$ lies in between, but closer to the 
 phenomenological ansatz.} 
 \label{fig:fig3}
\end{figure}

In Fig.~\ref{fig:fig3} we show the impact of resonance decays, focusing on the case $g_p=5$. We determine the 
final net-proton number cumulants including critical fluctuations via the expressions summarized in 
Appendix~\ref{sec:AppC} and consider the three different scenarios for the couplings $g_R$ between the critical 
mode and (anti-)resonances discussed in section~\ref{sec:2D}. We find that resonance decays influence the 
net-proton number cumulant ratios visibly, reducing the non-monotonicity induced by the critical fluctuations. 
Critical fluctuation signals are suppressed in $S\sigma$ by about $40$\% and in $\kappa\sigma^2$ by about $50$\%, 
i.e.~the reduction effect is slightly stronger for higher-order cumulants. This behavior is qualitatively in 
agreement with the predicted impact of isospin randomization processes on critical fluctuation 
signals~\cite{Kitazawa:2012at}. 
%%%%%%%%%%%%%%%%%%%%%%%%%%%%%%%%%%%%%%%%%%%%%%%%%%%%%%%%%%%%%%%%%%%%%%%%%%%%%%%%%%%%%%%%%%%%%%%%%%%%%%%%%%%%%%%%%%
%%%%%%%%%%%%%%%%%%%%%%%%%%%%%%%%%%%%%%%%%%%%%%%%%%%%%%%%%%%%%%%%%%%%%%%%%%%%%%%%%%%%%%%%%%%%%%%%%%%%%%%%%%%%%%%%%%

The results shown in Fig.~\ref{fig:fig3} indicate that resonance decays have significant effects on the cumulant 
ratios, but that for a sufficiently strong coupling $g_p$ between critical mode and (anti-)protons critical 
fluctuations survive final state processes such as resonance decays. Details in the couplings $g_R$ play a rather 
minor role as evident from Fig.~\ref{fig:fig3}. The strongest suppression of critical fluctuation signals is found 
for $g_R=0$, while the least is observed for $g_R$ in line with the phenomenological ansatz 
Eq.~(\ref{equ:sigmaresocoupling}). 

%%%%%%%%%%%%%%%%%%%%%%%%%%%%%%%%%%%%%
\subsection{Discussion\label{sec:3C}}
%%%%%%%%%%%%%%%%%%%%%%%%%%%%%%%%%%%%%

The results discussed in section~\ref{sec:3B} are obtained for fixed values of $g_p$. In general, it is reasonable 
to assume that the coupling strengths between the critical mode and (anti-)particles depend on $\sqrt{s}$. For 
example, in~\cite{Stephanov:1999zu} it was argued that at the critical point the $\sigma$-pion coupling could be 
significantly smaller than in vacuum. Moreover, quark-meson model~\cite{Abu-Shady:2015zpm} and Nambu--Jona-Lasinio 
(NJL) model~\cite{Hatsuda:1986gu} studies find that meson-nucleon couplings decrease both with increasing $T$ 
and/or $\mu_B$. Thus, by phenomenologically adjusting $g_p(\sqrt{s})$, as done in~\cite{Jiang:2015hri}, a 
quantitative model-to-data comparison would certainly be feasible, which is, however, beyond the scope of the 
present work. 

The specific behavior of the skewness, which shows a reduction of $S\sigma$ compared to the baseline result, 
depends on our choices for the direction of the $h$-axis relative to the $T$-axis, the relative sign of 
$\delta\sigma$ and $h$, and the sign of $g_p$. For simplicity we have taken all these signs to be positive. 
It is important to note that the sign of the critical contribution changes, leading to an enhancement of 
$S\sigma$ relative to the non-critical baseline, if either of these signs is changed. Indeed, in the case 
of the liquid-gas phase transition in ordinary liquids, the order parameter is $\delta n = n-n_{\textrm{cr}}$, 
which is mapped to the positive $h$-direction in the Ising model. This supports the idea of aligning $h$ with the 
positive $T$-axis in the QCD phase diagram, but it also suggests, based on Eq.~(\ref{equ:DeltaDist}) and $g>0$, 
that $\delta\sigma$ is aligned with the negative $h$-axis. The assumption $g>0$ is consistent with the simple 
sigma model picture that $g$ controls the dynamically generated mass, $m\sim g\sigma$. An NJL model 
study~\cite{Fujii:2004jt} of the $\sigma$-mode finds that in the critical region positive fluctuations in $\sigma$ 
are associated with positive fluctuations in the net-baryon density. In our simple phenomenological approach this 
corresponds to $g<0$, cf.~Eq.~(\ref{equ:DeltaDist}). A different NJL model study~\cite{Chen:2015dra} shows that 
the baryon skewness is negative on the high $(\mu_B,T)$ side of the transition line. Again, this result 
corresponds to changing one of the signs in our model. 

It is also important to quantify the influence of a critical point on the lowest-order cumulant ratio 
$\sigma^2/M$. According to Eq.~(\ref{equ:C2compact}), $\sigma^2/M$ is, compared to the non-critical baseline, 
increased through the critical fluctuation contributions in $C_2$. In contrast, neither the 
published~\cite{Adamczyk:2013dal} nor the preliminary STAR data~\cite{Thader:2016gpa} show within errors any 
sign of non-monotonic behavior in $\sigma^2/M$. This puts significant constraints on the values of $g_p$, 
$\Delta\mu_B^{\textrm{cp}}$, $\Delta T^{\textrm{cp}}$, $M_0$ and $H_0$, and the location of the critical point 
relative to the chemical freeze-out in the QCD phase diagram. 

\begin{figure}[tb]
 \includegraphics[width=0.47\textwidth]{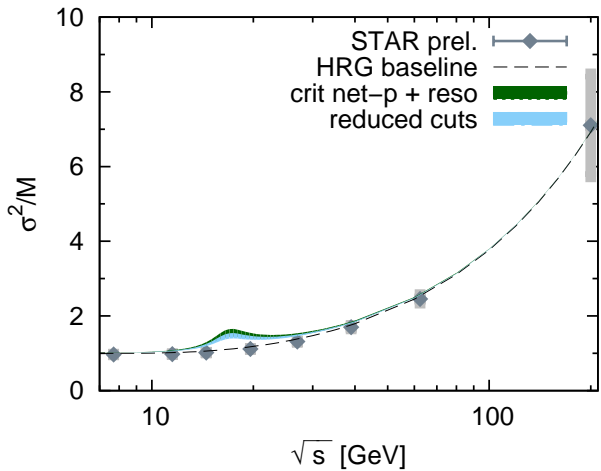}\\ \vspace{2mm}
 \caption{(Color online) Cumulant ratio $\sigma^2/M$ as a function of $\sqrt{s}$ including resonance decays and 
 critical fluctuation contributions for $g_p=5$ (both filled bands). As in Fig.~\ref{fig:fig3}, for $g_R=0$ the 
 results are closest to the non-critical baseline (thin dashed curve). For guidance, we also show the preliminary 
 data~\cite{Thader:2016gpa}. Reducing the kinematic acceptance, for example by decreasing the value for 
 $k_T^{\textrm{max}}$ from $2$~GeV (upper band) to $0.8$~GeV (lower band), reduces the non-monotonic behavior 
 caused by critical fluctuations slightly.} 
 \label{fig:fig4}
\end{figure}

In Fig.~\ref{fig:fig4}, we show $\sigma^2/M$ including the contributions from resonance decays for the case 
$g_p=5$ along the parametrized freeze-out curve. The filled bands comprise again the results of the three 
different ans\"atze for $g_R$, cf.~Fig.~\ref{fig:fig3}. For comparison also the preliminary STAR 
data~\cite{Thader:2016gpa} and the non-critical baseline (thin dashed curve) are shown. The non-monotonic 
structures seen in Fig.~\ref{fig:fig4} are caused by the critical fluctuations. By decreasing $k_T^{\textrm{max}}$ 
from $2$~GeV (upper band) to $0.8$~GeV (lower band), the non-monotonic behavior is only slightly reduced. The 
observed reduction is a consequence of the fact that the thermal factors $I_i$ in the cumulants correlate 
(anti-)particles of different momenta in the distribution with each other. By decreasing the considered kinematic 
acceptance, these correlations are reduced, as is also discussed e.g.~in~\cite{Ling:2015yau,Ohnishi:2016bdf}. The 
effect is found to be stronger for higher-order cumulant ratios. Moreover, increasing the considered rapidity 
range from $|y|\leq 0.5$ to $|y|\leq 0.75$ instead of changing $k_T^{\textrm{max}}$ from $0.8$~GeV to $2$~GeV has 
a quantitatively comparable influence, cf.~also~\cite{Ling:2015yau}. 
%%%%%%%%%%%%%%%%%%%%%%%%%%%%%%%%%%%%%%%%%%%%%%%%%%%%%%%%%%%%%%%%%%%%%%%%%%%%%%%%%%%%%%%%%%%%%%%%%%%%%%%%%%%%%%%%%%
%%%%%%%%%%%%%%%%%%%%%%%%%%%%%%%%%%%%%%%%%%%%%%%%%%%%%%%%%%%%%%%%%%%%%%%%%%%%%%%%%%%%%%%%%%%%%%%%%%%%%%%%%%%%%%%%%%

%%%%%%%%%%%%%%%%%%%%%%%%%%%%%%%%%%%%%%%%%%%%
\section{Conclusions\label{sec:Conclusions}}
%%%%%%%%%%%%%%%%%%%%%%%%%%%%%%%%%%%%%%%%%%%%

In this work, we have quantified the impact of resonance decays on higher-order net-proton number cumulant 
ratios for the case that the primordial net-proton number distribution contains critical fluctuations 
induced by the presence of a QCD critical point. The critical fluctuations were included on top of a 
non-critical background described by a thermal hadron resonance gas model. The coupling of particle and 
anti-particle resonances to the critical mode was modeled using a sigma model coupling. Fluctuations of 
the critical mode were determined, applying universality class arguments, from derivatives of the order 
parameter in the three-dimensional Ising model. We employed a linear map to relate QCD chemical freeze-out 
parameters $(\mu_B^{\textrm{fo}},T^{\textrm{fo}})$ to the Ising variables $(r,h)$. 

We find that resonance decays reduce, but do not completely obscure, the non-monotonic beam energy 
dependence of cumulant ratios induced by a critical point. The suppression is found to be slightly stronger for 
higher-order cumulants, in qualitative agreement with the influence of isospin randomization processes 
in the hadronic phase~\cite{Kitazawa:2012at}. This is fairly independent of the strength of the coupling between 
(anti-)resonances and the critical mode. Shrinking the kinematic acceptance reduces the critical point 
signals as expected. The coupling to the critical mode results in correlations between protons and 
anti-protons both from the direct coupling and from the indirect coupling via resonances. The correlations turn 
out to be fairly small, modifying the cumulant ratios by a few percent at most. 

Clearly, there are many issues that need to be addressed in order to make more quantitative model-to-data 
comparisons. This includes, in particular, the role of dynamical effects, both in the evolution of the 
order parameter and in the coupling to resonances. 

%%%%%%%%%%%%%%%%%%%%%%%%%%%
\section*{Acknowledgements}
%%%%%%%%%%%%%%%%%%%%%%%%%%%

The authors thank L.~Jiang, L.~Labun and K.~Redlich for useful discussions and F.~Geurts and J.~Th\"ader for 
providing the preliminary STAR data~\cite{Thader:2016gpa} shown in this work. M.B. thanks the Institute for 
Nuclear Theory at the University of Washington for the hospitality while this manuscript was finalized. M.N. 
acknowledges support from a fellowship within the Postdoc-Program of the German Academic Exchange Service (DAAD). 
This work was supported in parts by the U.S. Department of Energy under grants DE-FG02-03ER41260 and 
DE-FG02-05ER41367, and within the framework of the Beam Energy Scan Theory (BEST) Topical Collaboration. 

%%%%%%%%%
\appendix
%%%%%%%%%
\numberwithin{equation}{section}

%%%%%%%%%%%%%%%%%%%%%%%%%%%%%%%%%%%%%%%%%%%%%%%%%%%%%%%%%%%%%%%%%%%%%%%%%%%%%%%%%%%%%%%%%%
\section{Critical mode fluctuations based on universality class arguments\label{sec:AppA}}
%%%%%%%%%%%%%%%%%%%%%%%%%%%%%%%%%%%%%%%%%%%%%%%%%%%%%%%%%%%%%%%%%%%%%%%%%%%%%%%%%%%%%%%%%%

A suitable parametric representation of the magnetization $M(r,h)$ in the three-dimensional Ising model, which is 
valid in the scaling region around the critical point, is given in terms of the auxiliary variables $R$ and 
$\theta$ as~\cite{Guida:1996ep} 
\begin{align}
 \label{equ:A1}
 M & = M_0 R^\beta \theta \,,\\
 \label{equ:A2}
 r & = R (1-\theta^2) \,,\\
 \label{equ:A3}
 h & \equiv H/H_0 = R^{\beta\delta} \tilde{h}(\theta)
\end{align}
with $\tilde{h}(\theta)=c\theta\left(1+a\theta^2+b\theta^4\right)$ being an odd function of $\theta$ which is 
regular near $\theta=0$ and $\theta=\pm 1$. The parametric representation~\cite{Guida:1996ep} uses the universal 
critical exponents $\beta=0.3250$ and $\delta=4.8169$, while the coefficients in $\tilde{h}(\theta)$ read 
$a=-0.76145$, $b=0.00773$ and $c=1$ such that the relevant non-trivial roots of $\tilde{h}(\theta)$ are located at 
$\theta_0=\pm\sqrt{1.33128}$. Furthermore, the representation Eqs.~(\ref{equ:A1})~-~(\ref{equ:A3}) is uniquely 
defined for $R\geq 0$ and $-|\theta_0|\leq\theta\leq|\theta_0|$. For $R>0$ and $\theta=0$, we have $h=0$ and 
positive $r$, which corresponds to the crossover transition, while $\theta=\theta_0$ corresponds to the 
first-order phase transition line. The normalization constants $M_0$ and 
$H_0$~\cite{Schofield:1969zza,Josephson1969} are of mass dimension one and three, respectively, and the 
magnetization $M$ is positive for $(r<0,h\to 0^+)$ and for $(r\to 0,h>0)$. 

We obtain the equilibrium cumulants of the critical mode within this parametrization from 
Eq.~(\ref{equ:sigmacumulants}). They read as functions of $R$ and $\theta$ 
\begin{widetext}
 \begin{align}
%%%
  \label{equ:sigmaC2}
  \langle(\delta\sigma)^2\rangle & = \frac{T}{V} \frac{M_0}{cH_0} R^{\beta-\beta\delta} 
  \frac{\left(1+(2\beta-1)\theta^2\right)}{\left(1 + (3a+2\beta\delta -1)\theta^2 + 
  (a(2\beta\delta-3)+5b)\theta^4 + b(2\beta\delta-5)\theta^6\right)} \,, \\
%%%
  \label{equ:sigmaC3}
  \langle(\delta\sigma)^3\rangle & = - 2 \frac{T^2}{V^2} \frac{M_0}{(cH_0)^2} R^{\beta-2\beta\delta} 
  \frac{\theta \left(A_0+A_1\theta^2+A_2\theta^4+A_3\theta^6+A_4\theta^8\right)}{\left(1 + 
  (3a+2\beta\delta-1)\theta^2 + (a(2\beta\delta-3)+5b)\theta^4 + b(2\beta\delta-5)\theta^6\right)^3} 
  \,, \\
%%%
  \label{equ:sigmaC4}
  \langle(\delta\sigma)^4\rangle_c & = 2 \frac{T^3}{V^3} \frac{M_0}{(cH_0)^3} R^{\beta-3\beta\delta} 
  \frac{\left(B_0 + B_1\theta^2 + B_2\theta^4 + B_3\theta^6 + B_4\theta^8 + B_5\theta^{10} + B_6\theta^{12} 
  + B_7\theta^{14} + B_8\theta^{16}\right)}{\left(1 + (3a+2\beta\delta-1)\theta^2 + 
  (a(2\beta\delta-3)+5b)\theta^4 + b(2\beta\delta-5)\theta^6\right)^5} \,. 
 \end{align}
%\end{widetext}
The variance $\langle(\delta\sigma)^2\rangle$ is, by definition, a positive function. The coefficients in 
$\langle(\delta\sigma)^3\rangle$ are 
%\begin{widetext}
 \begin{align}
%%%
  A_0 = & \,\,3a+3\beta(\delta-1) \,,\\
%%%
  A_1 = & \,\,a\left[\beta(7\delta-3)-9\right]+2\beta(\delta-1)(\beta\delta+\beta-2)+10b \,,\\
%%%
  A_2 = & \,\,a\left[2\beta(\beta(\delta(\delta+4)-3)-6\delta)+9\right] 
  +\beta(2\beta-1)(\delta-1)(2\beta\delta-1)+b\left[11\beta\delta+5\beta-30\right] \,,\\
%%%
  A_3 = & \,\,a\left[(2\beta-1)(\beta(\delta-1)-1)(2\beta\delta-3)\right] 
  +2b\left[\beta^2(\delta(\delta+8)-5)-10\beta(\delta+1)+15\right] \,,\\
%%%
  A_4 = & \,\,b\left[(2\beta-1)(\beta(\delta-1)-2)(2\beta\delta-5)\right] \,,
 \end{align}
%\end{widetext}
and in $\langle(\delta\sigma)^4\rangle_c$ they read 
%\begin{widetext}
 \begin{align}
%%%
  B_0 = & \,\,-A_0 \,,\\
%%%
  B_1 = & \,\,45a^2+3a\left[2\beta(11\delta-7)+5\right]+12\beta^2(\delta-1)(3\delta-1)-30b \,,\\
%%%  
  \nonumber
  B_2 = & \,\,15a^2\left[11\beta\delta-3(\beta+5)\right]+6a\left[2\beta\left(17\beta\delta^2-3(6\beta+7)
  \delta+ 3\beta+17\right)-5\right]+255ab \\
  & +2\beta(\delta-1)(2\beta(\beta(\delta(11\delta-4)-1)-18\delta+6)+5)+30b\left[7\beta(\delta-1)+5\right] 
  \,,\\
%%%
  \nonumber
  B_3 = & \,\,30a^2(2\beta\delta-3)\left[\beta(4\delta-2)-5\right] \\
  \nonumber
  & +a\left[4\beta\left(82\delta+\beta\left(38\beta\delta^3-3\delta^2(6\beta+41)+6\delta(19-3\beta)
  +6\beta-9\right)-78\right)+30\right] \\
  \nonumber
  & +15ab\left[54\delta\beta-14\beta-85\right]
  +4\beta(\delta-1)(2\beta\delta-1)(\beta(2\beta(\delta(\delta+2)-1)-7\delta+3)+2) \\
  & +12b\left[2\beta(\delta(2\beta(8\delta-7)-29)+35)-25\right] + 450b^2 \,,\\
%%%  
  \nonumber
  B_4 = & \,\,6a^2\left[\beta(161\delta+2\beta(\delta(6-53\delta)+\beta(\delta(\delta(9\delta+11)-15)+3)
  +9)-33)-75\right] \\
  \nonumber
  & +a\left[4\beta\left(\beta\left(4\delta\beta^2(\delta(2\delta(\delta+5)-15)+4)-4\beta\left(16\delta^3
  -15\delta+3\right)+\delta(97\delta-74)-9\right)-41\delta+45\right)-15\right] \\
  \nonumber
  & +6ab\left[2\beta(\beta(\delta(67\delta+2)-15)-263\delta+55)+425\right]
  +\beta(2\beta-1)(\delta-1)(1-2\beta\delta)^2(\beta(4\delta-2)-1) \\
  \nonumber
  & +4b\left[2\beta^3\left(3\delta\left(9\delta^2+3\delta-11\right)+5\right)+6\beta^2(\delta(34-41\delta)
  +5)+5\beta(39\delta-59)+75\right] \\
  & +15b^2\left[\beta(59\delta+5)-150\right] \,,\\
%%%
  \nonumber
  B_5 = & \,\,a^2(2\beta\delta-3)\left[2\beta\left(4\beta^2(\delta(\delta(\delta+9)-12)+3)
  -6\beta(\delta(7\delta+6)-6)+77\delta-6\right)-75\right] \\
  \nonumber
  & +a\left[(2\beta-1)(2\beta\delta-3)(2\beta\delta-1)\left(2\beta\left(\beta\left(4\delta^2-6\delta
  +2\right)-5\delta+4\right)-1\right)\right] \\
  \nonumber
  & +2ab\left[2\beta\left(2\beta^2(\delta(\delta(35\delta+87)-93)+15)-9\beta(\delta(61\delta+26)-25)
  +8(142\delta-15)\right)-1275\right] \\
  \nonumber 
  & +2b\big[4\beta\big(85+2\beta\big(2\beta^2\delta^4+6\beta\delta^3(3\beta-4)+\delta^2(52-3\beta(8\beta
  +7)) \\
  \nonumber
  & \quad\qquad+6\delta(\beta(\beta+7)-7)-5(\beta+3)\big)-39\delta\big)-75\big] \\
  & +12b^2\big[\beta\left(53\beta\delta^2+10\delta(4\beta-29)-25(\beta+2)\right)+375\big] \,,\\
%%%
  \nonumber
  B_6 = & \,\,a^2\left[(2\beta-1)(\beta(\delta-1)-1)(\beta(4\delta-2)-5)(3-2\beta\delta)^2\right] \\
  \nonumber
  & +ab\big[4\beta(\beta(4\delta\beta^2(\delta(2\delta(\delta+13)-33)+8) - 4\beta(\delta(2\delta(16\delta
  +51)-105)+15) \\
  \nonumber
  & \quad\qquad+\delta(499\delta+434)-315)-5(143\delta+9))+1275\big] \\
  \nonumber
  & +2b\big[(2\beta-1)(2\beta\delta-5)(2\beta\delta-1)\left(\beta\left(\beta\left(4\delta^2-
  6\delta+2\right)-9\delta+7\right)-3\right)\big] \\
  & +2b^2\left[\beta\left(\beta^2(2\delta(\delta(43\delta+177)-165)+50)-
  24\beta(\delta(37\delta+45)-25)+2545\delta+775\right)-2250\right] \,,\\
%%%
  \nonumber
  B_7 = & \,\,ab(2\beta\delta-5)\left[(2\beta-1)(2\beta\delta-3)\left(2\beta\left(\beta\left(4\delta^2-
  6\delta+2\right)-13\delta+10\right)+17\right)\right] \\
  & +b^2(2\beta\delta-5)\left[4\beta\left(2\beta^2(\delta(\delta(\delta+17)-21)+5)-5\beta(\delta(7\delta
  +22)-15)+119\delta+80\right)-450\right] \,,\\
%%%
  B_8 = & \,\,b^2\left[(2\beta-1)(\beta(\delta-1)-2)(5-2\beta\delta)^2(\beta(4\delta-2)-9)\right] \,.
 \end{align}
\end{widetext}
In the limit of the so called linear parametric representation with critical exponents $\beta=1/3$ and $\delta=5$ 
and coefficients $a=-2/3$, $b=0$ and $c=3$ these expressions simplify to the cumulants reported 
in~\cite{Stephanov:2011pb,Mukherjee:2015swa}. 

One can study the dependence of the cumulants of the critical mode on the spin model variables by analyzing, for 
example, the dimensionless skewness 
$\tilde{S}=\langle(\delta\sigma)^3\rangle/\langle(\delta\sigma)^2\rangle^{3/2}$ and kurtosis 
$\tilde{\kappa}=\langle(\delta\sigma)^4\rangle_c/\langle(\delta\sigma)^2\rangle^2$, which both diverge at the 
critical point. For this purpose, we use Eqs.~(\ref{equ:A2}) and~(\ref{equ:A3}) to convert given values for $r$ 
and $h$ into $R$ and $\theta$. The skewness is an odd function of $h$, 
cf.~also~\cite{Chen:2016sxn,Mukherjee:2015swa}, which is positive for $h<0$ and changes sign continuously 
(discontinuously) for positive (negative) $r$ when $h$ becomes positive. The kurtosis, instead, is an even 
function of $h$, which for a given $r>0$ is negative within an interval $|h|<h^*(r)$ which depends on $r$. For 
$r<0$, one finds positive $\tilde{\kappa}$ as function of $h$ with a cusp at $h=0$. As a function of $r$, 
$\tilde{S}$ exhibits a maximum (minimum) for negative (positive) $h$ at an $r>0$ which itself depends on $h$, 
while $\tilde{\kappa}$ exhibits with decreasing positive $r$ first a minimum followed by a maximum whose locations 
depend on the non-zero value of $|h|$. 

Let us compare the parametric expressions for the cumulants of the critical mode in 
Eqs.~(\ref{equ:sigmaC2})~-~(\ref{equ:sigmaC4}) with those obtained from the corresponding Ginzburg-Landau 
effective potential. Including cubic and quartic interactions one finds~\cite{Mukherjee:2015swa,Stephanov:2008qz} 
\begin{align}
 \label{equ:sigma2}
 \langle (\delta\sigma)^2 \rangle & = \frac{T}{V}\xi^2 \,,\\
 \label{equ:sigma3}
 \langle (\delta\sigma)^3 \rangle & = -2\tilde{\lambda}_3\frac{T^{3/2}}{V^2}\xi^{9/2} \,,\\
 \label{equ:sigma4}
 \langle (\delta\sigma)^4 \rangle_c & = 6\left(2\tilde{\lambda}_3^2-\tilde{\lambda}_4\right)\frac{T^2}{V^3} 
 \xi^7 \,,
\end{align}
where the correlation length dependence of the cubic and quartic coupling strengths as known in the scaling region 
has already been taken into account, such that $\tilde{\lambda}_3$ and $\tilde{\lambda}_4$ are dimensionless 
parameters. 

\begin{figure}[tb]
 \includegraphics[width=0.47\textwidth]{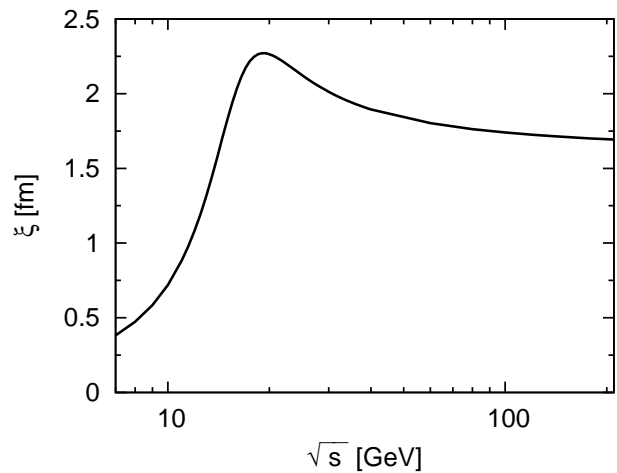}\\ \vspace{2mm}
 \caption{Beam energy dependence of $\xi$ along the considered chemical freeze-out curve (see 
 Fig.~\ref{fig:fig1}). The correlation length falls off slower in the crossover regime, i.e.~at larger $\sqrt{s}$, 
 as is expected from the underlying universality class. The depicted result will be described qualitatively with 
 the phenomenological ansatz for $\xi$ introduced in~\cite{Athanasiou:2010kw}, if a comparable size of the 
 critical region is considered.} 
 \label{fig:fig5}
\end{figure}

Comparing $\langle(\delta\sigma)^2\rangle$ in Eq.~(\ref{equ:sigmaC2}) with Eq.~(\ref{equ:sigma2}) allows us to 
determine the $\sqrt{s}$-dependence of $\xi$ along the chemical freeze-out curve considered in this work. This is 
shown in Fig.~\ref{fig:fig5}. 
%%%%%%%%%%%%%%%%%%%%%%%%%%%%%%%%%%%%%%%%%%%%%%%%%%%%%%%%%%%%%%%%%%%%%%%%%%%%%%%%%%%%%%%%%%%%%%%%%%%%%%%%%%%%%%%%%%
%%%%%%%%%%%%%%%%%%%%%%%%%%%%%%%%%%%%%%%%%%%%%%%%%%%%%%%%%%%%%%%%%%%%%%%%%%%%%%%%%%%%%%%%%%%%%%%%%%%%%%%%%%%%%%%%%%
One finds that the correlation length falls off slower in the crossover regime of the QCD phase diagram in line 
with the underlying universality class. Details of this behavior depend very sensitively on the chosen values for 
the normalization constants $M_0$ and $H_0$, but also on the details of the mapping such as the size of the 
critical region and the proximity of the chemical freeze-out to the QCD phase transition. For the parameter 
choices specified in section~\ref{sec:3B} we find that $\xi$ becomes as large as $2.3$~fm for 
$\sqrt{s}\simeq 19.2$~GeV, which corresponds to a freeze-out baryon chemical potential $\mu_B^{\textrm{fo}}$ 
significantly smaller than $\mu_B^{\textrm{cp}}$. A similar comparison of the higher-order cumulants can be made 
to determine the behavior of the dimensionless parameters $\tilde{\lambda}_3$ and $\tilde{\lambda}_4$. These are 
found to be independent of $R$ and, thus, remain finite in the domain $-|\theta_0|\leq\theta\leq|\theta_0|$ of 
the parametric representation. In particular, $\tilde{\lambda}_3$ turns out to be negative for $h<0$. This implies 
that the sign of the skewness $\tilde{S}$ depends sensitively on the probed values of $h$. 

%%%%%%%%%%%%%%%%%%%%%%%%%%%%%%%%%%%%%%%%%%%%%%%%%%%%%%%%%%%%%%%%%%%%%%%%%%
\section{Parametrization of the chemical freeze-out curve\label{sec:AppB}}
%%%%%%%%%%%%%%%%%%%%%%%%%%%%%%%%%%%%%%%%%%%%%%%%%%%%%%%%%%%%%%%%%%%%%%%%%%

The chemical freeze-out conditions reported in~\cite{Alba:2014eba} were determined by analyzing the published STAR 
data~\cite{Adamczyk:2013dal,Adamczyk:2014fia} of $\sigma^2/M$ for event-by-event net-proton number and 
net-electric charge distributions. The qualitative behavior of these observables is dominated by charged pions and 
(anti-)protons. The published results for net-proton number fluctuations~\cite{Adamczyk:2013dal} were obtained by 
considering a significantly smaller kinematic acceptance in transverse momentum $k_T$ compared 
to~\cite{Luo:2015ewa,Luo:2015doi,Thader:2016gpa}. Within errors, no pronounced deviation from non-critical 
baseline expectations was found in the data~\cite{Adamczyk:2013dal,Adamczyk:2014fia} for the lowest-order cumulant 
ratios. This justified the application of the non-critical baseline model in the analysis~\cite{Alba:2014eba}. 

In particular for larger $\sqrt{s}$, the chemical freeze-out temperatures found in~\cite{Alba:2014eba} are 
significantly lower than those obtained in traditional approaches in which yields or yield ratios are analyzed, 
cf.~e.g.~\cite{Cleymans:2005xv}. Moreover, for small $\mu_B$ a positive slope of the chemical freeze-out curve is 
found, cf.~Fig.~\ref{fig:fig1}. This is opposite to the curvature of the crossover line obtained in lattice QCD, 
as can be seen from Eq.~(\ref{equ:crossoverline}) and the cited values for $\kappa_c$. Nonetheless, both small 
values for the chemical freeze-out temperature~\cite{Borsanyi:2014ewa,Bazavov:2015zja} and a positive slope of the 
freeze-out curve for small $\mu_B$~\cite{Bazavov:2015zja} have also been obtained in lattice QCD analyses of the 
same fluctuation observables~\cite{Adamczyk:2013dal,Adamczyk:2014fia}. 

A practical parametrization of the freeze-out conditions~\cite{Alba:2014eba} can be given in form of functions of 
the beam energy $\sqrt{s}$. We define analytic functions for the chemical potentials at chemical freeze-out 
$\mu_X^{\textrm{fo}}$ similar to~\cite{Cleymans:2005xv} reading 
\begin{equation}
 \label{equ:AppC1}
 \mu_X^{\textrm{fo}}(\sqrt{s}) = \frac{d_X}{\left(e_X\sqrt{s}+f_X\right)^{g_X}} \,,
\end{equation}
which yield $\mu_X^{\textrm{fo}}$ in GeV given $\sqrt{s}$ in GeV for the parameter values listed in 
Tab.~\ref{tab:tab1}. The chemical freeze-out temperature $T^{\textrm{fo}}$ is best parametrized as a function of 
$\mu_B^{\textrm{fo}}$ via 
\begin{equation}
 \label{equ:AppC2}
 T^{\textrm{fo}}(\mu_B^{\textrm{fo}}) = t_0 + t_1\mu_B^{\textrm{fo}} + t_2\left(\mu_B^{\textrm{fo}}\right)^2 
 + t_4\left(\mu_B^{\textrm{fo}}\right)^4 + t_6\left(\mu_B^{\textrm{fo}}\right)^6 
\end{equation}
with parameter values $t_0=0.146$~GeV, $t_1=0.079$, $t_2=-0.366$~GeV$^{-1}$, $t_4=0.251$~GeV$^{-3}$ and 
$t_6=-0.107$~GeV$^{-5}$ yielding $T^{\textrm{fo}}$ in GeV for $\mu_B^{\textrm{fo}}$ in GeV. The fit is shown in 
Fig.~\ref{fig:fig1} by the solid curve. These parametrizations give reasonable descriptions of the freeze-out 
conditions~\cite{Alba:2014eba} and hold for beam energies $\sqrt{s}\geq 7$~GeV. The non-vanishing values for 
$\mu_Q$ and $\mu_S$ for a given $\sqrt{s}$ are a consequence of physical side conditions for first cumulants, 
namely overall strangeness neutrality and an initial proton to baryon ratio of about $0.4$ as realized in 
Au$+$Au or Pb$+$Pb heavy-ion collisions. Due to the smallness of $\mu_B^{\textrm{fo}}$ at high beam energies, the 
latter condition is still quantitatively appropriate at high $\sqrt{s}$ even though the created system is almost 
isospin symmetric at mid-rapidity. 
%%%%%%%%%%%%%%%%%%%%%%%%%%%%%%%%%%%%%%%%%%%%%%%%%%%%%%%%%%%%%%%%%%%%%%%%%%%%%%%%%%%%%%%%%%%%%%%%%%%%%%%%%%%%%%%%%%
\begin{table}[htb]
{
\setlength{\extrarowheight}{1.9pt}
\begin{tabular}{|c|c|c|c|c|}
\hline
$X$  &  $d_X/$GeV    & $e_X/$GeV$^{-1}$ &  $f_X$  & $g_X$ \\ \hline\hline 
B    &    \,\,1.161      &       0.392      & \,-0.481  & 0.910 \\ \hline
Q    &   -0.386      &       2.822      & 12.319  & 1.070 \\ \hline
S    &    \,\,0.848      &       1.138      &  \,\,\,1.297  & 0.995 \\ \hline
\end{tabular}
\caption{\label{tab:tab1}
Parameter values for the parametrizations of $\mu_X^{\textrm{fo}}$ as functions of $\sqrt{s}$ in 
Eq.~(\ref{equ:AppC1}).}
}
\end{table}
%%%%%%%%%%%%%%%%%%%%%%%%%%%%%%%%%%%%%%%%%%%%%%%%%%%%%%%%%%%%%%%%%%%%%%%%%%%%%%%%%%%%%%%%%%%%%%%%%%%%%%%%%%%%%%%%%%

%%%%%%%%%%%%%%%%%%%%%%%%%%%%%%%%%%%%%%%%%%%%%%%%%%%%%%%%%%%%%%%%%%%%%%%%%%%%%%%%%%%%%%%%%%%%%%%%%%%%%%%%%%%%%%%
\section{Resonance decay contributions to net-proton cumulants including critical fluctuations\label{sec:AppC}}
%%%%%%%%%%%%%%%%%%%%%%%%%%%%%%%%%%%%%%%%%%%%%%%%%%%%%%%%%%%%%%%%%%%%%%%%%%%%%%%%%%%%%%%%%%%%%%%%%%%%%%%%%%%%%%%

The variance of the final net-proton number is given by $\hat{C}_2=\langle (\Delta \hat{N}_p)^2 \rangle + 
\langle (\Delta \hat{N}_{\bar{p}})^2 \rangle - 2 \langle\Delta \hat{N}_p\Delta \hat{N}_{\bar{p}} \rangle$, where 
the two-particle correlators including resonance decays read~\cite{Fu:2013gga,Begun:2006jf} 
\begin{align}
\nonumber
 \langle (\Delta\hat{N}_i)^2\rangle = & \,\, \langle (\Delta N_i)^2\rangle + 
 \sum_R \langle (\Delta N_R)^2\rangle \langle n_i\rangle_R^2 \\ 
 \label{equ:C2critReso}
 & + \sum_R \langle N_R \rangle \langle (\Delta n_i)^2 \rangle_R \,,\\ 
 \label{equ:C11critReso}
 \langle \Delta\hat{N}_i\Delta\hat{N}_j\rangle = & \,\, \langle \Delta N_i\Delta N_j\rangle + 
 \sum_R \langle \Delta N_R\Delta N_{\bar{R}}\rangle \langle n_i\rangle_R \langle n_j\rangle_{\bar{R}}
\end{align}
with $\langle(\Delta n_i)^2\rangle_R=\sum_r b_r^R (n_{i,r}^R)^2 - \langle n_i\rangle_R^2$. Note again that the 
sums over $R$ include all resonances and anti-resonances of the non-critical baseline model such that if $R$ 
denotes an anti-resonance like $\bar{\Lambda}(1520)$, the corresponding $\bar{R}$ is the resonance 
$\Lambda(1520)$. 

According to Eq.~(\ref{equ:C11critReso}), resonance decays can induce correlations between different particle 
species. Nonetheless, without critical fluctuations protons and anti-protons would remain uncorrelated in our 
grandcanonical hadron resonance gas model approach because of baryon charge conservation in each decay. The 
individual contributions $\langle (\Delta N_R)^2\rangle$ and $\langle \Delta N_R\Delta N_{\bar{R}}\rangle$ in 
Eqs.~(\ref{equ:C2critReso}) and~(\ref{equ:C11critReso}), which include critical fluctuations, follow from 
Eqs.~(\ref{equ:C2crit}) and~(\ref{equ:C11crit}) analogous to $\langle (\Delta N_i)^2\rangle$ and 
$\langle \Delta N_i\Delta N_j\rangle$ for (anti-)protons. Critical point induced correlations between a resonance 
and its anti-resonance are, therefore, inherited by their decay products. This induces additional correlations 
between protons and anti-protons on top of the primordial correlations discussed in section~\ref{sec:2B}. 

From Eqs.~(\ref{equ:C2critReso}) and~(\ref{equ:C11critReso}) we find that $\hat{C}_2$ contains non-critical 
fluctuation contributions, which are independent of the correlation length, and critical fluctuation contributions 
of order ${\cal O}(\xi^2)$. This implies that the critical contribution to the resonance terms scales with 
the same power of $\xi$ as in the primordial result Eq.~(\ref{equ:C2compact}). The situation is more complicated 
in the case of the cumulants of third and fourth order, see~\cite{Fu:2013gga}. Consider the third-order cumulant. 
The critical contribution to the primordial term, Eq.~(\ref{equ:C3crit}), scales as $\xi^{9/2}$. The resonance 
terms contain a contribution, induced by fluctuations $\langle (\Delta N_R)^3\rangle$, that scales with the same 
power of $\xi$, but they also contain contributions, induced by the probabilistic nature of branching, that scale 
with lower powers of $\xi$. Since we do not keep subdominant terms, we drop these contributions in the resonance 
terms. The same statement applies to the fourth-order cumulant. 

For $\hat{C}_3 = \langle(\Delta \hat{N}_p)^3\rangle - \langle(\Delta \hat{N}_{\bar{p}})^3\rangle 
- 3 \langle(\Delta \hat{N}_p)^2\Delta \hat{N}_{\bar{p}}\rangle + 3 \langle\Delta \hat{N}_p(\Delta 
\hat{N}_{\bar{p}})^2\rangle$ the three-particle correlators including resonance decays were derived 
in~\cite{Fu:2013gga}. Taking only non-critical fluctuation contributions and critical fluctuation contributions of 
order ${\cal O}(\xi^{9/2})$ into account, we find for these correlators 
\begin{widetext}
 \begin{align}
  \label{equ:C3critResoFinal}
  \langle(\Delta \hat{N}_i)^3 \rangle = & \,\, \langle(\Delta N_i)^3 \rangle + 
  \sum_R \langle (\Delta N_R)^3\rangle \langle n_i\rangle_R^3 + 
  3 \sum_R C_2^R \langle(\Delta n_i)^2\rangle_R \langle n_i\rangle_R + 
  \sum_R \langle N_R\rangle\langle(\Delta n_i)^3\rangle_R \,,\\
  \label{equ:C21critResoFinal}
  \langle(\Delta \hat{N}_i)^2 \Delta \hat{N}_j \rangle = & \,\, \langle(\Delta N_i)^2 \Delta N_j \rangle + 
  \sum_R \langle (\Delta N_R)^2 \Delta N_{\bar{R}} \rangle \langle n_i\rangle_R^2 \langle n_j\rangle_{\bar{R}} 
  \,,
 \end{align}
\end{widetext}
where $\langle(\Delta n_i)^3\rangle_R = \sum_r b_r^R (n_{i,r}^R)^3 - 3 \langle n_i\rangle_R \sum_r b_r^R 
(n_{i,r}^R)^2 + 2 \langle n_i\rangle_R^3$. The individual contributions in Eq.~(\ref{equ:C3critResoFinal}) follow 
from Eqs.~(\ref{equ:CnDefinition}) and~(\ref{equ:C3crit}) and in Eq.~(\ref{equ:C21critResoFinal}) from 
Eq.~(\ref{equ:C21crit}). 

For $\hat{C}_4 = \langle(\Delta \hat{N}_p)^4 \rangle_c + \langle(\Delta \hat{N}_{\bar{p}})^4 \rangle_c 
- 4 \langle(\Delta \hat{N}_p)^3\Delta \hat{N}_{\bar{p}} \rangle_c 
+ 6 \langle(\Delta \hat{N}_p)^2(\Delta \hat{N}_{\bar{p}})^2 \rangle_c 
- 4 \langle\Delta \hat{N}_p(\Delta \hat{N}_{\bar{p}})^3 \rangle_c$ the individual fourth-order cumulants follow 
from the four-particle correlators including resonance decays, cf.~\cite{Fu:2013gga}. Taking for $\hat{C}_4$ only 
non-critical fluctuation contributions and critical fluctuation contributions of order ${\cal O}(\xi^7)$ into 
account, we find 
\begin{widetext}
 \begin{align}
  \nonumber
  \langle(\Delta \hat{N}_i)^4 \rangle_c = & \,\, \langle(\Delta N_i)^4 \rangle_c + 
  \sum_R \langle (\Delta N_R)^4\rangle_c \langle n_i\rangle_R^4 + 
  6 \sum_R C_3^R \langle(\Delta n_i)^2\rangle_R \langle n_i\rangle_R^2 \\
  \label{equ:C4critResoFinal}
  & \,\, + \sum_R C_2^R \left[3 \langle(\Delta n_i)^2\rangle_R^2 + 4 \langle(\Delta n_i)^3\rangle_R 
  \langle n_i\rangle_R \right] + \sum_R \langle N_R\rangle \langle(\Delta n_i)^4\rangle_{R,c} \,,\\
  \label{equ:C22critResoFinal}
  \langle(\Delta \hat{N}_i)^m(\Delta \hat{N}_j)^n \rangle_c = & \,\, \langle(\Delta N_i)^m(\Delta N_j)^n 
  \rangle_c + \sum_R \langle(\Delta N_R)^m(\Delta N_{\bar{R}})^n \rangle_c \langle n_i\rangle_R^m 
  \langle n_j\rangle_{\bar{R}}^n 
 \end{align}
\end{widetext}
for $m+n=4$, where $\langle(\Delta n_i)^4\rangle_{R,c} = \langle(\Delta n_i)^4\rangle_R - 
3 \langle(\Delta n_i)^2\rangle_R^2$ with 
$\langle(\Delta n_i)^4\rangle_R=\sum_r b_r^R (n_{i,r}^R)^4 - 4 \langle n_i\rangle_R \sum_r b_r^R 
(n_{i,r}^R)^3 + 6 \langle n_i\rangle_R^2 \sum_r b_r^R (n_{i,r}^R)^2 - 3 \langle n_i\rangle_R^4$. The individual 
contributions in Eq.~(\ref{equ:C4critResoFinal}) follow from Eqs.~(\ref{equ:CnDefinition}) 
and~(\ref{equ:C4crit}) and in Eq.~(\ref{equ:C22critResoFinal}) from Eq.~(\ref{equ:Cmncrit}). 

%%%%%%%%%%%%%%%%%%%%%%%%%%%

%%%%%%%%%%%%%%%%%%%%%

\end{document}